\documentclass{POS}
\usepackage{amssymb}
\usepackage{times}
\usepackage{graphicx}

\title{The ridge laboratory}

\author{\speaker{Thomas Peitzmann}\\
        Utrecht University, Utrecht, The Netherlands\\
        E-mail: \email{t.peitzmann@uu.nl}}

\abstract{Mechanisms proposed to explain the jet-related correlation of long range in rapidity (\textit{the ridge}) observed in high-energy heavy-ion collisions are reviewed. Limitations of a model using the combined effect of transverse flow and a direction bias of partons suffering energy loss are discussed. The influence of the time scale of the correlation mechanism on the rapidity range is investigated.}

\FullConference{4th international workshop High-pT physics at LHC 09\\
		 February 4-7, 2009\\
		 Prague, Czech Republic}

\ShortTitle{The ridge laboratory}

\begin{document}

\section{Introduction}
One of the most interesting results obtained in measurements of heavy-ion collisions at RHIC is the existence of the so-called ``ridge'' \cite{star-ridge}. For a trigger hadron of relatively high transverse momentum, the angular correlation with associated hadrons on the same side in central heavy-ion collisions can be described as a two-component structure:
\begin{itemize}
\item a narrow peak symmetric in $\Delta \phi$ and $\Delta \eta$ similar in strength and shape to correlation structures in p+p collisions interpreted as originating from jets and
\item an enhancement narrow in $\Delta \phi$ but broad in $\Delta \eta$ (the ridge). 
\end{itemize}

\begin{figure}[b]
\begin{center}
\includegraphics[width=0.8\textwidth]{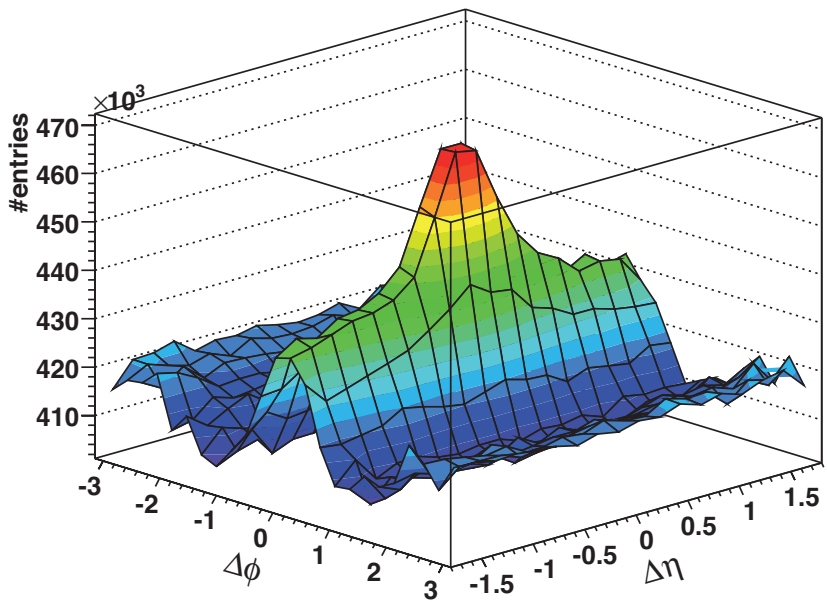}
\caption{Yield of particles associated with a high $p_T$ trigger hadron as a function of $\Delta \phi$ and $\Delta \eta$ in central Au+Au collisions at 200 GeV/nucleon as measured by the STAR experiment \cite{star-ridge}.}
\label{fig:ridge}
\end{center}
\end{figure}

An example of a correlation function as measured by the STAR experiment is shown in Fig.~\ref{fig:ridge}. Findings related to the ridge phenomenon are:
\begin{enumerate}
\item It persists up to very large momenta of the trigger particle ($p_T \approx 8 \, \mathrm{GeV}/c$), suggesting a jet-related origin.
\item The shape shows a very weak decrease with increasing $\Delta \eta$ and is consistent with a constant for $\Delta \eta < 1.8$. This large range in pseudorapidity requires the correlation to be generated at very early times, when the different rapidity regions are still causally connected.
\item The momentum distribution of the ridge-associated particles is very similar to that of the bulk of produced particles (or underlying event) and as such different from the distribution of particles associated with the jet-like peak.
\end{enumerate}
There are observations of similar correlation structures for particles of lower momentum, which do not appear to allow a similar, simple separation into a jet-peak and a ridge \cite{star-minijet}. It is still under discussion  how closely these two phenomena may be related. In what follows I will mostly relate to the properties of the former (hard ridge) and not so much discuss possible explanations for the latter (soft ridge). Also, recently the PHOBOS experiment has reported a semi-soft ridge to exist still for a pseudorapidity separation of  $\Delta \eta \approx 4$ \cite{phobos-ridge}.

\section{Mechanisms for ridge generation}
The phenomenology of the ridge suggests that there exists a correlation in $\Delta \phi$ between a hadron from jet fragmentation with a part of the bulk matter spread out in rapidity. Different mechanism leading to the ridge have been discussed in the literature, but none has so far been clearly identified.

In \cite{voloshin-ridge1} it is first discussed qualitative that transverse flow effects may lead to correlations in rapidity, but no quantitative estimates are given.
According to \cite{dumitru-ridge, gavin-ridge2} quantum fluctuations extending over a large range in rapidity from glasma flux tubes may be responsible. Here superimposed effects of transverse flow lead to angular correlations. \cite{gavin-ridge} discusses in particular the interplay of viscosity and transverse flow. These investigations  \cite{dumitru-ridge, gavin-ridge2,gavin-ridge} are intended as an explanation for the ``soft ridge'', and will likely not explain the structure associated with a high $p_T$ trigger particle.
In \cite{ma-ridge} attempts are made to explain the ridge as longitudinal broadening of jets. The obtained width is however too narrow to explain the observations.

\begin{figure}
\includegraphics[width=\textwidth]{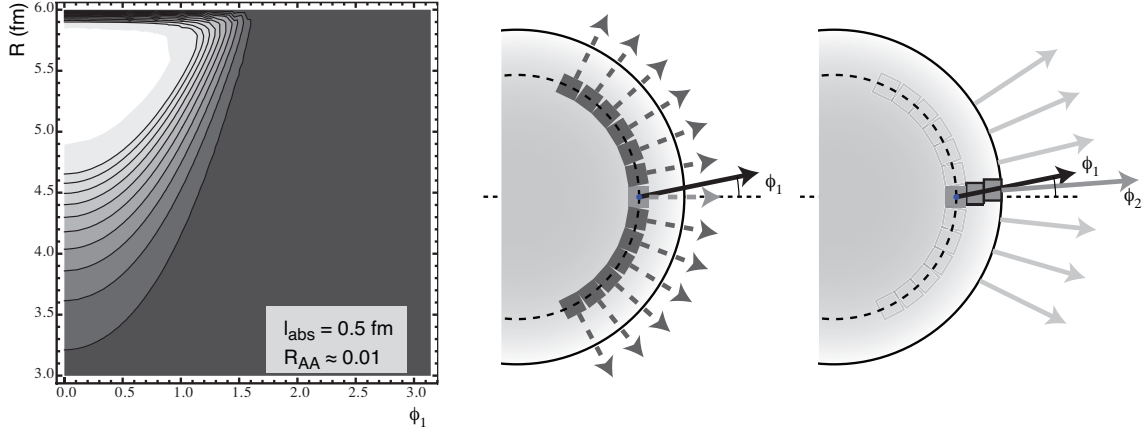}
\caption{Left: Distribution of number of triggers from hard scatterings as a function of their radial position $R$ and emission angle $\phi_1$ relative to the surface normal following the prescription of \cite{shuryak-ridge}. Also indicated is the expected inclusive suppression factor $R_{AA}$. Center and right: Illustration of angular emission patterns of jet and bulk as discussed in the text.}
\label{fig1}
\end{figure}

Ref. \cite{shuryak-ridge} introduces a simple model to estimate the correlations arising from the simultaneous effects of parton energy loss and transverse flow. It relies on the fact that from energy loss the hard-scattered parton providing the trigger particle has a directional bias. The spatial position of the hard scattering will be close to the surface and the direction of emission will be focused along the outgoing radial direction. Simple parameterizations for the distribution of the hard scattering points $P_{prod}$ assuming collision scaling (eq. 2 in the paper) and the quenching probability $P_{quench}$ assuming a simple exponential damping with a characteristic quenching length $l_{abs}$ (eq. 3) are given. The product $P_{trig}(r,\phi_1) \equiv P_{prod} \cdot P_{quench}$ then describes the distribution of source points and emission angles of the observable trigger particles. Figure~\ref{fig1} shows on the left the distributions for a typical quenching length ($l_{abs} = 0.5 \, \mathrm{fm}$) studied in \cite{shuryak-ridge}.\footnote{In the presentation at the workshop erroneous estimates of the inclusive suppression have been used - this is corrected here.} The source points are very strongly biased to the surface and the partons are strongly focussed in emission angle in this case. As mentioned in the paper, this focussing is the origin of the azimuthal angle correlation of the ridge structure in this model. It is argued in the paper that bulk matter (originating from flux tubes, strings etc.) produced around the point of hard scattering shows a similar direction bias as the trigger from radial flow. The results given in \cite{shuryak-ridge} show still a too large angular width, which already calls for additional mechanisms to reproduce the data. Moreover, the absorption length of $l_{abs} = 0.5 \, \mathrm{fm}$ leads to a much stronger inclusive suppression ($R_{AA} \approx 0.01$) than observed experimentally. More realistic values of $R_{AA}$ are obtained with absorption lengths of the order of $l_{abs} \approx 3 \, \mathrm{fm}$. For such parameter values the angular correlation would be much weaker, as is mentioned in \cite{shuryak-ridge}. 
In addition, bulk matter elements at all possible emission angles $\phi_2$ should have similar yield of particles as illustrated in Figure~\ref{fig1} in the center, while a visible correlation structure in $\Delta \phi = \phi_1 - \phi_2$ as claimed in \cite{shuryak-ridge} would need a higher yield at small relative angles. In fact, the hard scattering will take away some energy from the volume element in question, which would rather translate to a reduced yield from the bulk at small relative angles. This makes it unlikely that this model by itself can explain the ridge correlation. 

However, there is a natural mechanism, which can enhance the bulk matter. The energy loss of the parton is likely to be deposited in the bulk system leading to a boost or thermal enhancement and in consequence to a larger yield at the same angle visible as a correlation structure (see Figure~\ref{fig1} right).

Such an interaction of jets with the medium is used by another class of models, e.g. via momentum kicks (i.e. elastic energy loss) \cite{wong-ridge} or via gluon radiation enhancing the thermal medium \cite{hwa-ridge}. We will not discuss those mechanisms in detail here, but both models seem to explain many aspects of the experimental phenomena. If these, or similar explanations are valid this would make the ridge a very useful tool to study the jet-medium interaction. There is, however, a limit in the possible rapidity range for such model where the correlation is created by final state interactions, as I will discuss below.

In \cite{voloshin-ridge2} the authors estimate effects of collective flow on jets. Quantitatively this leads to interesting results, but the major assumption of the calculations, maximal coupling of jets to collective flow is certainly questionable. It is very briefly discussed how jets may obtain e.g. transverse flow. One of the possible mechanisms given, initial state $k_T$ broadening, does not lead to \emph{collective} flow -- so no space-momentum correlations. Another mechanism mentioned relates to radial colour fields. Those should be delayed relative to longitudinal fields \cite{fries-color} and should thus only act on partons after the hard scattering. While the fragmentation of a jet is certainly modified by such final state effects (like e.g. parton energy loss), it is unlikely that it will so strongly influence the higher $p_T$ fragments, which appear to emerge like in vacuum fragmentation. 

A large extent in rapidity can naturally be explained by initial state effects. Mechanisms have to provide some coupling between a hard scattered parton and the underlying event. Brodsky has proposed \cite{brodsky} that the directional bias in initial $k_T$ of the parton one gets from using a high $p_T$ trigger should also be reflected in a similar bias in the DGLAP radiation of that parton. The latter radiation would be distributed over a broad range in rapidity, possibly interact with the matter present and turn into hadrons leading to the ridge. As for all such initial state effects one would expect a direct photon trigger to be accompanied by a similar correlation, while for final state mechanisms photons should have a strongly reduced correlation.

In principle, Fermi motion of the nucleons could introduce a similar effect. Here a directional bias of the hard scattered partons would also be carried by the spectator partons in the same nucleon, which should contribute to the bulk matter. However, the small magnitude of Fermi momentum may be insufficient to produce a strong enough effect. 

\section{Time scales and rapidity range}
A limit on the time when the correlation can be last introduced is given in \cite{dumitru-ridge}. For two particles freezing out at $\tau_{\mathrm{fo}}$ to be causally related with a rapidity separation $2 \Delta y$, the process responsible has to happen at a time:
\begin{equation}
\tau < \tau_{\mathrm{fo}} \cdot \exp\left(- \Delta y \right).
\label{eq1}
\end{equation}
For the symmetric case of two (soft) particles both emitted from the same mechanism as discussed in \cite{dumitru-ridge} $\Delta y$ corresponds to half the rapidity difference between the two. If one deals with a hard parton defining the rapidity of the trigger particle and emitting another particle (e.g. a gluon) which then creates the associated hadron $\Delta y$ would just be the rapidity difference between those.\footnote{In the first case both particles can move longitudinally with respect to their common origin, while in the second the associated particle can move longitudinally relative to the trigger.} 
This limit does, however, not take into account transverse degrees of freedom. In particular mechanisms involving energy transfer from a high-energy parton to the bulk will involve particles (e.g. radiated gluons) carrying transverse momentum. If their longitudinal velocity is just a fraction $\beta_L < 1$, the correlation cannot stretch out as far in rapidity. An analog estimate to eq. \ref{eq1} yields:
\begin{equation}
\tau < \tau_{\mathrm{fo}} \cdot \left[ \exp\left(- \Delta y \right) - \frac{1-\beta_L}{\beta_L} \sinh\left(\Delta y \right) \right].
\label{eq2}
\end{equation}
The requirement for final state effects $\tau>0$ then yields a limited rapidity range, which is just the rapidity of the emitted gluon in the longitudinal rest frame of the hard scattered parton:
\begin{equation}
\Delta y_{max} \approx \frac{1}{2} \ln \left( \frac{1 + \beta_L}{1-\beta_L} \right).
\label{eq3}
\end{equation}
For an emission angle of e.g. $60^{\circ}$ the maximum rapidity range of the correlation would already only be of the order  of $\Delta y_{max} \approx 1.3$. As gluons reaching to the higher rapidities would have less transverse momentum, one should in addition expect a decrease of the strength of the correlation with increasing rapidity difference. The STAR data do not completely rule out such a dependence -- a careful analysis of this dependence will be very interesting. But it is difficult to explain a correlation at $\Delta y \approx 4$ as observed by PHOBOS  \cite{phobos-ridge} with such mechanisms.

\section{Conclusions}

There exists at present no theoretical model quantitatively explaining all experimental observations related to the ridge. Models using final state interactions must run into the constraints of causality, which will limit the rapidity range and should also lead to a decrease of the correlation strength for larger rapidity difference. However, if these can be shown to be responsible for the effect (or some part of it), the ridge would provide an interesting laboratory to study both parton energy loss and properties of the bulk matter. The strength of the correlation should be a measure of the amount of energy loss. Furthermore, the associated particles in the ridge represent a sample of the bulk from a relatively well defined region of the fireball with probably slightly higher temperature and/or velocity. So all studies of the thermal properties -- certainly hadrochemical composition and inverse slopes for different species, more speculatively also photon and charm production -- of this matter may be studied in a controlled way.

Models involving initial state correlations are free from the limitations imposed by interaction time scales, so they could extend over the entire rapidity range. Such models still have to be quantified and compared to experimental data. 

A number of additional tests should be performed:
\begin{itemize}
\item For initial state mechanisms one would expect the ridge structure to be also present for photon triggers, while it should be strongly reduced in this case for models built on parton energy loss. 
\item Dijet events should have different sensitivity for the structure -- for final state mechanisms because of their different surface bias and for initial state mechanisms because of the approximate momentum balance.
\item Heavy quark triggers should also show a difference for final state models. A slightly weaker correlation would be expected. May be the dead-cone effect in gluon radiation would even lead to a more complicated angular substructure of the correlation.
\end{itemize}

For LHC this will certainly remain an interesting topic. The larger accessible rapidity range should make the discrimination of initial and final state effects clearer.
The larger background from the increased multiplicity will make such studies more difficult, but at the same time, the much larger dynamic range  in $p_T$ and higher yield for hard scattered partons will be advantageous. 

\noindent{\small \textit{ I'd like to thank the organisers for the stimulating workshop. Fruitful discussions with S.~Brodsky and T.~Renk are gratefully acknowledged.}}


\begin{thebibliography}{99}
\bibitem{star-ridge} J.~Putschke et al. (STAR Collaboration), \emph{J.Phys.G} {\bf 34} (2007) S679; [{\tt arXiv:nucl-ex/0701074}].
\bibitem{star-minijet} J.~Adams et al. (STAR Collaboration), \emph{J.Phys.G} {\bf 32} (2006) L37.
\bibitem{phobos-ridge} E.~Wenger et al. (PHOBOS Collaboration), 
\emph{J.Phys.G} {\bf 35} (2008) 104080; [{\tt arXiv:0804.3038}].
\bibitem{voloshin-ridge1} S.A.~Voloshin, \emph{Phys.Lett. B} {\bf 632} (2006) 490; [{\tt arXiv:nucl-th/0312065}].
\bibitem{dumitru-ridge} A.~Dumitru et al., \emph{Nucl.Phys.A} {\bf 810} (2008) 91; [{\tt arXiv:0804.3858}].
\bibitem{gavin-ridge2} S.~Gavin, L. McLerran, and G. Moschelli (2008) [{\tt arXiv:0806.4366}].
\bibitem{gavin-ridge} S.~Gavin and G. Moschelli, \emph{J.Phys.G} {\bf 35} (2008) 104084; [{\tt arXiv:0806.4366}].
\bibitem{ma-ridge} G.L.~Ma et al., \emph{Eur.Phys.J.C} {\bf 57} (2008) 589; [{\tt arXiv:0807.3987}].
\bibitem{shuryak-ridge} E.~Shuryak, \emph{Phys. Rev. C} {\bf 76} (2007) 047901; [{\tt arXiv:0706.3531}].
\bibitem{wong-ridge} C.Y.~Wong (2009) [{\tt arXiv:0901.0726}].
\bibitem{hwa-ridge} C.B.~Chiu and R.C.~Hwa (2009) [{\tt arXiv:0809.3018}].
\bibitem{voloshin-ridge2} C.A.~Pruneau, S.~Gavin, and S.A.~Voloshin, \emph{Nucl.Phys.A} {\bf 802} (2008) 107; [{\tt arXiv:0711.1991}].
\bibitem{fries-color} R.J.~Fries, J.I.~Kapusta, and Y.~Li (2006)  [{\tt arXiv:nucl-th/0604054}].
\bibitem{brodsky} S.~Brodsky, these proceedings.
\end{thebibliography}
\end{document}